**Exact solution of the three-dimensional (3D) $Z_2$ lattice gauge theory**

Zhidong Zhang

Shenyang National Laboratory for Materials Science, Institute of Metal Research,

Chinese Academy of Sciences, 72 Wenhua Road, Shenyang, 110016, P.R. China

**Abstract**

The gauge invariance is an important phenomenon widely occurring in various matters with many-body interactions. The gauge theories are very important for understanding the evolution of fundamental physical processes in particle physics, high-energy physics and condensed matter physics. In this work, the origin of nonlocal effects is inspected and the contributions of nontrivial topological structures to physical properties are investigated in details for both the 3D Ising model and the 3D $Z_2$ lattice gauge model. Then the exact solution for the 3D $Z_2$ lattice gauge theory is derived by the duality between the two models. The partition function, the spontaneous magnetization, the true range of the correlation and the critical point are rigorously derived for the 3D $Z_2$ lattice gauge theory. The critical exponents of the 3D $Z_2$ lattice gauge theory are in the same universality class as the 3D Ising model, which are $\alpha = 0$, $\beta = 3/8$, $\gamma = 5/4$, $\delta = 13/3$, $\eta = 1/8$ and $\nu = 2/3$. Several fundamental issues, such as dimensionality, duality, symmetry, manifold, degenerate states, are investigated for these many-body interacting spin systems. The connections with superfluid, superconductors, particle physics, etc. are evaluated. Furthermore, physical significances and mathematical aspects of the 3D $Z_2$ lattice gauge theory are discussed with respect to topology, geometry, and algebra. The present results for the 3D $Z_2$ lattice gauge theory are helpful for figuring out the basic features of 3D lattice gauge theories with other symmetries, such as U(1), SU(2) and SU(3). This work shines a light on the interdisciplinarity between many-body interacting systems, algebra, topology and geometry.

The corresponding author: Z.D. Zhang, e-mail address: zdzhang@imr.ac.cn

## 1.Introduction

Phase transitions occur in almost every physical system, which have attracted great interest of physicists. Specifically, the continuous (second-order) phase transitions in condensed matters (such as, magnets, superconductors, superfluids, etc.) implicate abundant physical phenomena. The study of these continuous phase transitions and the critical phenomena at a critical point reveals the nature of interactions between spins (or particles). Because the models (for instance, the Ising model) for describing a magnet or a superconductor are very basic, they can uncover the fundamental characters of many-body interacting spin (or particle) systems, thus can be applicable for understanding the nature in other physical systems. In particular, the duality relationship may exist between some models that are utilized to interpret the physics in different fields as diverse as condensed matter physics, particle physics and high-energy physics.

Non-Abelian gauge theories [1,2] have become the focus of widespread interest, owing to their central role in understanding the evolution of fundamental physical processes: the quantum flavor dynamics of electroweak interactions and the quantum chromodynamics of strong interactions. The non-Abelian gauge theories were invented by Yang and Mills about seventy years ago [1]. The well-known Glashow-Weinberg-Salam model unifies electromagnetic with weak interactions, where gauge fields are identified with massless photons, and with the hypothetical massive vector mesons mediating weak interactions [3-8]. In the electroweak theory, a key point is that the mass was introduced with the Higgs mechanism by spontaneous symmetry breaking [9-12]. For the theory of strong interactions, it follows a fact that the forces in the dynamical model are asymptotic freedom, becoming negligible at short distances [13-15]. Consequently, Yang-Mills SU(3) fields coupled to quarks (quantum chromodynamics) provided the only realistic framework, by which experiments on high-energy lepton-nucleon scattering can be accommodated. It is a fact (up to date) that the model containing non-Abelian Yang-Mills gauge fields is the only theory in which this particular behavior appears possible [16]. Of course, the Yang-Mills field equations have not been solved exactly, not even in the context of classical field theory [17].

The lattice gauge theories can be utilized to imitate some behaviors of the Yang-Mills gauge theories, however, in the limit of the lattice space. Wegner invented Ising lattice gauge theory in 1971 [18]. The $Z_2$ lattice gauge theory is the simplest one among the lattice gauge theories [18-22], since it possesses the simplest symmetry $Z_2$, compared with other lattice gauge theories with U(1), SU(2) and SU(3) symmetries. We shall be interested in the large-distance properties of a $Z_2$ gauge theory assuming that the effective coupling is sufficiently large so that we can use Wilson's strong-coupling methods [19]. An ultraviolet cutoff is introduced into the field theory through a spatial lattice. Most of the space-time symmetries of relativistic field theories are destroyed by this construction, so the theory discussed here is not a realistic Yang-Mills theory. However, following Wilson [19], we are mainly interested in determining the special effects of exact gauge invariance in strongly coupled gauge theories. Keeping these in mind, what we are sure is that the exact solution of the three-dimensional (3D) $Z_2$ lattice gauge theory is extremely important for physicists not only for understanding the behaviors of this theory itself, but also for figuring out the basic features of other 3D lattice gauge theories with U(1), SU(2) and SU(3) symmetries [18-24]. To date, no exact solution for the 3D $Z_2$ lattice gauge theory has been reported yet.

Duality was illustrated between the 3D $Z_2$ lattice gauge theory and the 3D Ising model [18-24]. The Ising model is one of the simplest physical models describing many-body spin (or particle) interactions [25-31]. In the previous work [27-29], two conjectures were proposed by the present author in [27] and then proven rigorously in collaboration with Suzuki and March [29] for solving analytically the ferromagnetic 3D Ising model at the zero external magnetic field.

The motivations of the present work are as follows: The exact solution of the 3D $Z_2$ lattice gauge theory is a hard unsolved problem in physics, which is in the same difficulty level as the exact solution of the ferromagnetic 3D Ising model. The exact solution of the 3D Ising model provides an opportunity to determine the exact solution of the 3D $Z_2$ lattice gauge theory. The exact solution is extremely important for inspecting the fundamental laws for many-body interacting systems, and it can give a guide on numerical studies of these systems, which have been very rare. Furthermore, the features revealed by the exact solution for the 3D $Z_2$ lattice gauge theory provide an implication on the features of the 3D U(1), SU(2) and SU(3) lattice gauge theories, serving as a starting point for solving analytically these much more complicated models.

The remainder of this article is arranged along the following line of presentation: In Section 2, we inspect the origin of nonlocal effects in the 3D Ising models in details.

In Section 3, we explore the consequences of the obtained solutions in the dual formulation to derive the exact solutions for the physical properties (including partition function, critical point, spontaneous magnetization, spin correlation, critical exponents, etc.) of the 3D $Z_2$ lattice gauge theory. In Section 4, the exact solutions for the critical point and the critical exponents of the 3D Ising model are compared with other approximation methods. In Section 5, we discuss the physical significances and the mathematical aspects of the solutions. The results obtained in the 3D Ising model and the 3D $Z_2$ lattice gauge theory are connected with superfluid, superconductors, particle physics, with new thoughts on dimensionality, duality, symmetry, manifold, degenerate states, also with respect to topology, geometry, and algebra. Section 6 is for conclusions.

## 2.Nonlocal effects and nontrivial topological structures in the 3D Ising models

In the previous work [27-29,32-37], we uncovered the nonlocal effects and nontrivial topological structures in the 3D Ising models. In this section, we inspect the origin of the nonlocal effect in details. Specially we focus our attentions on contributions of nonlocal effects and nontrivial topological structures to the physical properties in the 3D Ising models.

In the quantum statistics mechanism, transfer matrices are used to represent the partition function of a physical system, to describe the probability of finding the system in a given configuration. The elements of the transfer matrices are determined by calculations of energies for all configurations of spins located at every lattice point in the system. As revealed in our previous work [27-29,32-37], the nonlocal effects in the 3D Ising model originate from the contradictory between the 3D character of the lattice and the 2D character of the transfer matrices used in the quantum statistics mechanism.

At first, let us start from the origin of the nontrivial topological structures. In a 3D Ising model, Ising spins are assigned on every lattice point of a 3D lattice with the lattice size $N = mnl$. The numbers $(i, r, s)$ running from (1, 1, 1) to $(m, n, l)$ denote lattice points along three crystallographic directions in the 3D lattice. One may denote the lattice points, a layer by a layer, by the number $j = [mn(s-1)+ m(r-1) + i]$, which runs in a sequence as 1, 2, 3,... , $mnl$. Such two representations are equivalent. The size of the transfer matrices in spinor representation for the 3D Ising model is $2^N \times 2^N$. It should be emphasized that the sequence in a process of a layer by a layer is the simplest one for mapping a 3D lattice into a 2D "lattice" (a matrix). For a study of the 3D Ising model, this sequence can remain some basic characters of the 2D Ising model, and make the procedure for solving the 3D Ising model as simple as possible. It is understood that any other sequences will make the problems much more complicated.

Let us inspect the running of $j$. For the first layer ($s$ =1), corresponding to the running from (1,1,1) to $(m, n, 1)$, we have $j$ = 1,2,3,…, $m$ for the first line ($r$ =1), $j$ = $m$+1, $m$+2, $m$+3,…, $2m$ for the second line ($r$ =2), …, and $j = (n-1)m+1,(n-1)m+2,…,$ $mn$ for the last line ($r = n$). For the second layer ($s$ =2), corresponding to (1, 1, 2) to $(m, n, 2)$, $j$ runs from $(mn+1)$, $(mn+2)$, $(mn+3)$,…, $2mn$. It runs in all the way to the last layer ($s = l$), $j$ runs from $(mn(l-1)+1)$, $(mn(l-1)+2)$, $(mn(l-1)+3)$,…, $mnl$. The first spin (1,1,1) in the first layer and the first spin (1, 1, 2) in the second layer correspond to j = 1 and j = $(mn+1)$, respectively. Figure 1 illustrates an Ising model on a 3D lattice with the size of 3×3×3, for an example, which is mapped into the spin arrangement on a 2D lattice with the size of (3×3+3×3+3×3), as arranged in the transfer matrix. Green, purple, blue colors represent the interactions along three crystallographic directions (although their values are equal for the simple cubic lattice, the colors are used for a clear illustration). In the transfer matrix, the interaction with blue color show the crossings. Clearly, the interaction between the two nearest neighboring spins behaves as a long range interaction, which involves the entanglements of all the spins in a plane. It is emphasized here that for solving the solution of the 3D Ising model, the system is in the thermodynamic limit ($m \to \infty$, $n \to \infty$, $l \to \infty$). This is the origin of the nontrivial topological structures in the 3D Ising models.

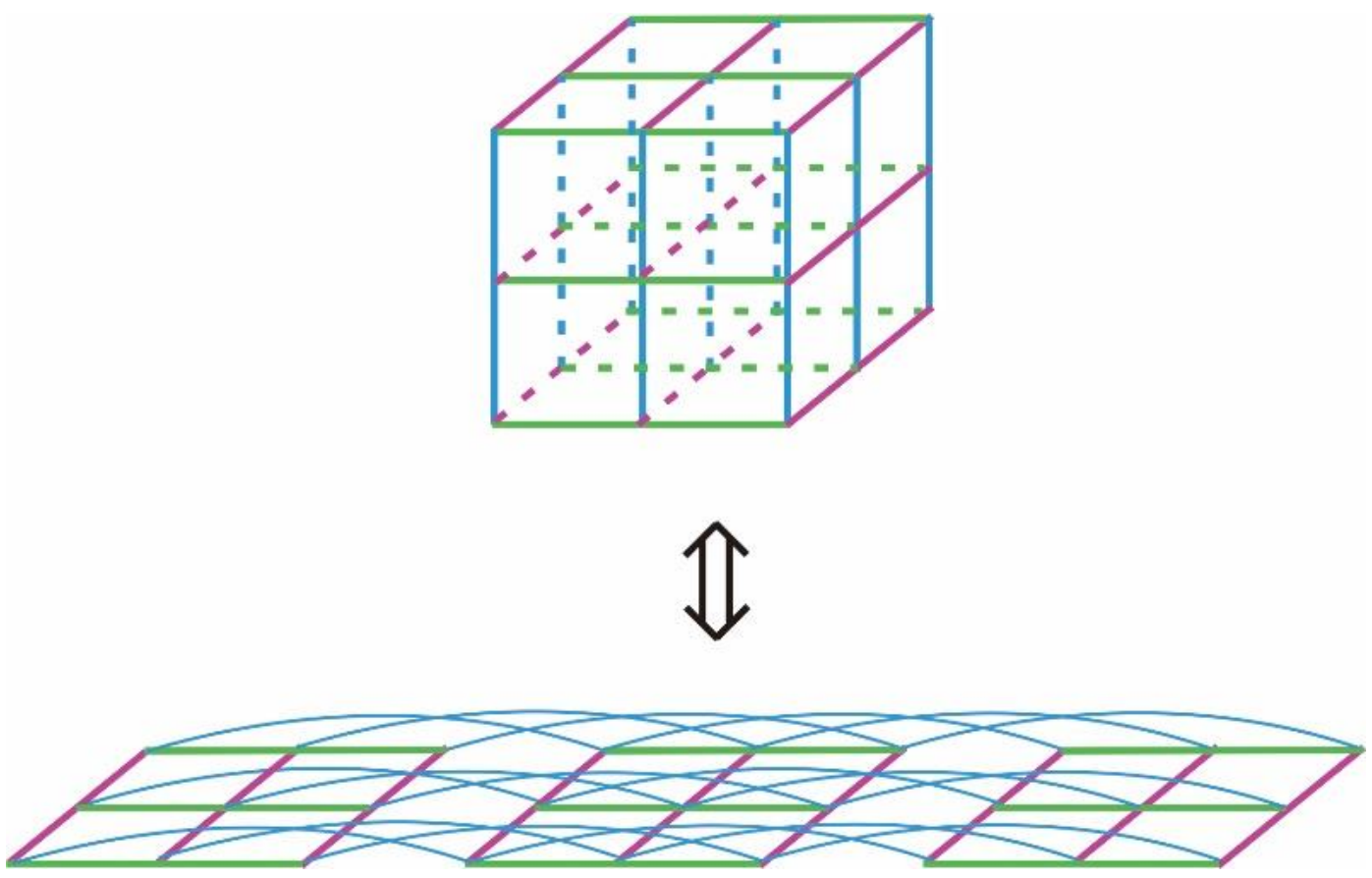

Figure 1. Illustration of a 3D Ising model on a 3×3×3 lattice, for an example, which is mapped into the 2D spin arrangement on a (3×3+3×3+3×3) lattice, as in the transfer matrix. Green, purple, blue colors represent the interactions along three crystallographic directions. In the transfer matrix, the interaction with blue color shows the crossings, indicating the existence of non-trivial topological structures in the 3D Ising model.

For simplicity, we can apply the cylindrical crystal model preferred by Onsager [26]

and Kaufman [38], in which we wrap our crystal around cylinders. However, unlike in the solid energy band theory for one-electron approximation in which the periodic boundary conditions can be applied along three crystallographic directions, in the present many-body interaction system, we can perform the periodic boundary condition only along one crystallographic direction. After performing the periodic boundary condition, the running number $j$ can be reduced to $j = [(s\text{-}1)n + r]$, running as $j = 1,2,3,\ldots,nl$ in a plane. The size of the transfer matrices in spinor representation for the 3D Ising model is reduced to be $2^{nl}\times 2^{nl}$. However, this simplicity does not alter the fact that in the 3D Ising model, the nonlocal effects, viewed as the long-range spin entanglements or the nontrivial topological structures, exist.

According to the topological theory [39-42], one can find that the Kauffman bracket polynomial is identical to the partition function of an Ising model. A mapping exists between the states of the crossings and the Ising spin alignments (spin up and spin down) with values of +1 and -1 [33], so that one finds the equivalence between a 2D knot and a 2D Ising model [39], and the equivalence between a braid and a spin chain [33]. The mapping between the Ising spin lattice and the knot structure can be generalized to the 3D case [33]. In the 3D Ising model (and the 3D $Z_2$ lattice gauge theory), two contributions to the partition functions exist: 1) the local alignments of spins and 2) the global effects of long-range spin entanglements. The 3D Ising model can be described as spin alignments at the 3D lattice plus braids along the third spatial direction. During the procedures for deriving the solution of the 3D Ising models, we reveal some characteristics of the systems, which are helpful for setting up novel quantum statistics mechanics, which accounts the contributions of the nontrivial topological structures [43].

In order to deal with the nontrivial topological structure in the 3D Ising model, we have introduced a rotation with an angle $K''' = K'K''/K$ in an additional dimension for the local gauge transformation for the simple orthorhombic Ising lattices [27-29]. The critical exponents of the 3D Ising models are $\alpha = 0$, $\beta = 3/8$, $\gamma = 5/4$, $\delta = 13/3$, $\eta = 1/8$ and $\nu = 2/3$ [27]. In ref. [44], experimental data were compared with these theoretical results to show that the 3D Ising universality indeed exists for critical indices in a certain class of magnets and at fluid–fluid phase transition. The exact solution for the critical exponents of the 3D Ising model agree with experimental data in various materials [44-48]. It is worth noticing that in a recent Monte Carlo simulation [49,50] the critical exponents of the 3D Ising model obtained by taking into account the long-range interactions of spin chains (namely, the nontrivial topological contribution) agree well with our exact solution.

## 3 Exact solution of a 3D $Z_2$ lattice gauge theory

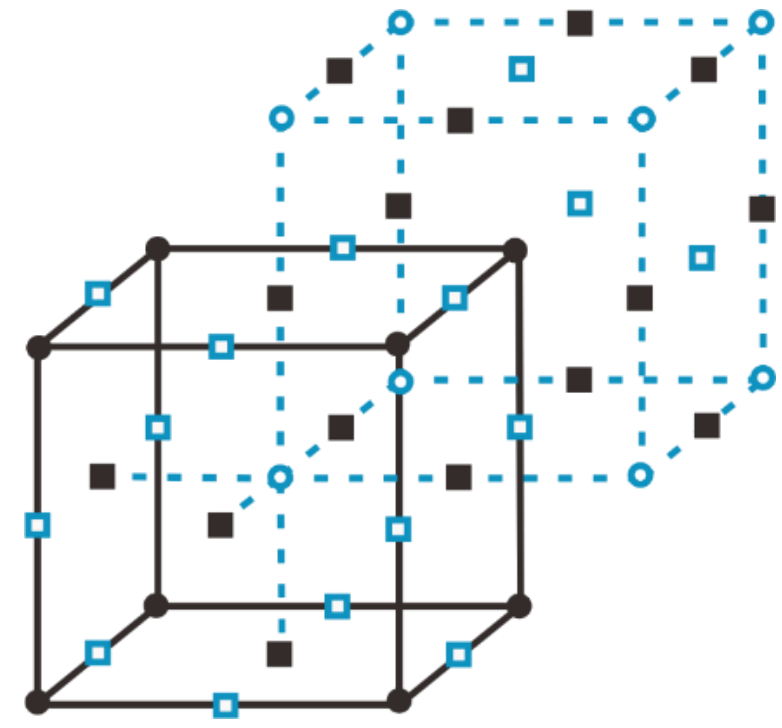

Figure 2 A simple cubic lattice (black solid lines) and its dual lattice (blue dashed lines) [18]. The black solid circles denote the corners $r^{(0)} = (i, r, s)$ of the original lattice, while the blue open circles denote the corners $r^{(3)} = (\mathrm{i} + 1/2, r + 1/2, s + 1/2)$ of the dual lattice. The blue open squares mark the intersect points $r^{(1)}$ at which the edges (black solid lines) of the original lattice and the faces of the dual lattice, while the black solid squares mark the intersect points $r^{(2)}$ at which the faces of the original lattice and the edges (blue

broken lines) of the dual lattice.

For studying the 3D $Z_2$ lattice gauge theory, let us consider a simple cubic lattice in d-dimensional ($d = 3$) Euclidean space. Label links of the lattice by a site $n$ and a unit lattice vector $\mu$ (or $\nu$). Notice that the same link can be labeled as $(n, \mu)$ or $(n + \mu, -\mu)$. Place Ising spins ($\sigma_3 = \pm 1$) on links. The duality between two cubic lattices was illustrated in Figure 2 (see also Figure 3 of ref. [18]), where a cube of the original cubic lattice ($d = 3$) and a cube of the dual lattice were drawn. The corners $r^{(0)} = (i, r, s)$ of the original lattice are denoted by black solid circles, while the corners $r^{(3)} = (i + 1/2, r + 1/2, s + 1/2)$ by blue open circles. Interestingly, the edges (continuous lines) of the original lattice and the faces of the dual lattice intersect at points $r^{(1)}$ (blue open squares), whereas the faces of the original lattice and the edges (broken lines) of the dual lattice intersect at points $r^{(2)}$ (black solid squares).

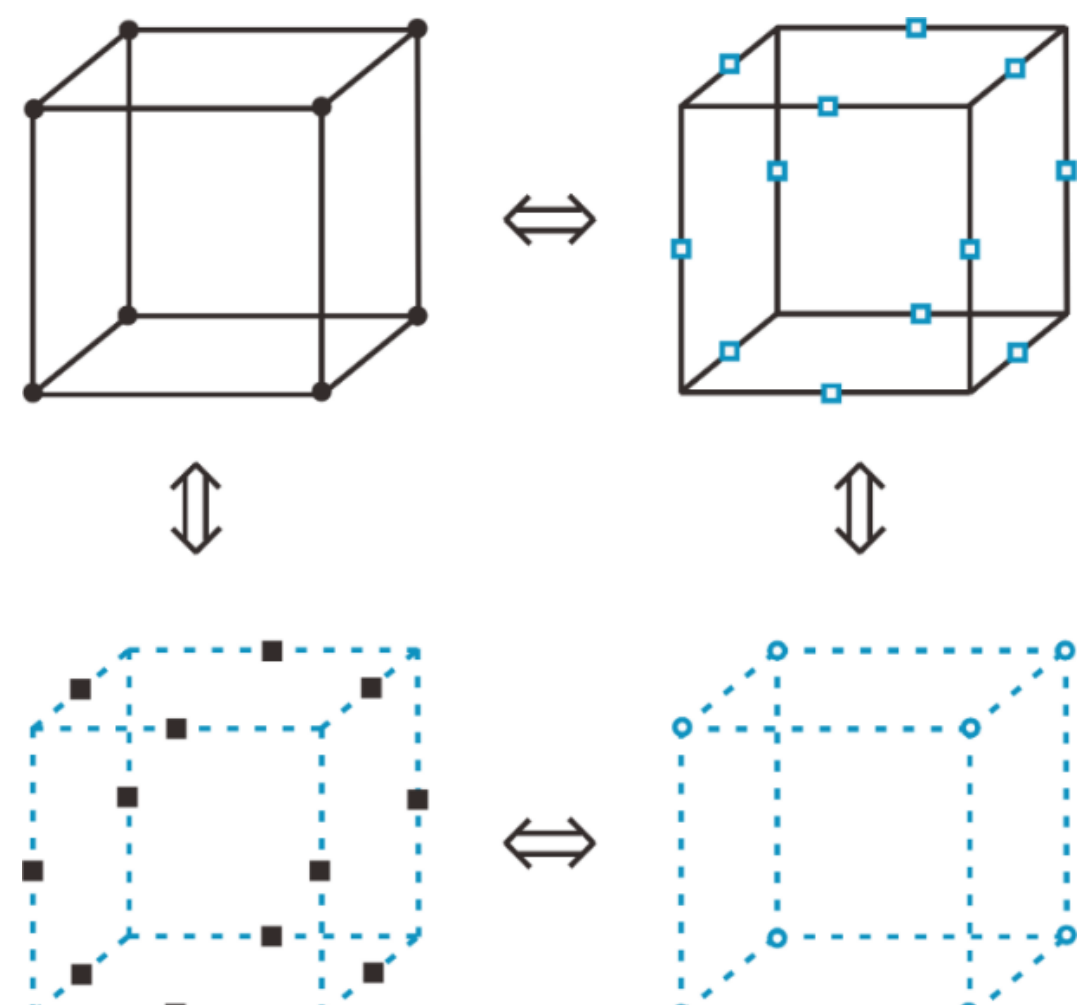

Figure 3. Mapping between a simple cubic lattice (black solid lines) and its dual lattice (blue dashed lines). There are four models, which can be mapped each other: A 3D Ising model with interaction K on the original lattice (left on top), a 3D $Z_2$ lattice gauge

model with interactions $K^*$ on the original lattice (right on top), a 3D Ising model with interactions $K^*$ on the dual lattice (right on bottom) and a 3D $Z_2$ lattice gauge model with interaction K on the dual lattice (left on bottom). The spins in the 3D Ising models are located at the corners of the lattices, while the spins in the 3D $Z_2$ lattice gauge model are located at the edges of the lattices.

The mapping between a cube of the original cubic lattice ($d = 3$) and a cube of the dual lattice is illustrated in Figure 3. There are four models, which can be mapped each other in a cycle of mappings. The 3D $Z_2$ lattice gauge theory with interaction $J^*$ on a cubic lattice is described by the Action (see Eq.(5.7) of [21]):

$$S = -J^* \sum_{n,\mu\nu} \sigma_3(n,\mu)\sigma_3(n+\mu,\nu)\sigma_3(n+\mu+\nu,-\mu)\,\sigma_3(n+\nu,-\nu) \tag{1}$$

Note that $J^*$ accounts for the interaction on the product of Ising spins around plaquettes (or primitive squares) of the lattice. The partition function of the 3D $Z_2$ lattice gauge theory is represented as (see also Eq. (2.26) or (2.27) of ref. [24]):

$$Z = 2^{-N/2}(sinh2K^*)^{-3N/2} \sum_{\{A\}} exp\left( K^* \sum_{n,\mu\nu} \sigma_3(n,\mu)\sigma_3(n+\mu,\nu)\sigma_3(n+\mu+\nu,-\mu)\sigma_3\,(n+\nu,-\nu) \right) \tag{2}$$

where $K^* = \frac{J^*}{k_B T}$, with the Boltzmann constant $k_B$ and the temperature $T$. A local gauge transformation at the site n can be defined as the operation G(n) of flipping all the spins on links connected to that site. Because G(n) can be applied anywhere, the Action has a huge invariance group. A nontrivial action with this symmetry consists of the product of spins around plaquettes of the lattice. Clearly, an Ising spin model can be constructed with this local symmetry. As illustrated in details in [21,24], a mapping exists between the high- (low-) temperature properties of the 3D Ising gauge system and the low- (high-) temperature properties of the 3D Ising spin system. The partition function of the 3D Ising model with Ising spins ($s_i = \pm 1$) is written as (see Eq. (2.20) or (2.31) of ref. [24]):

$$Z = \sum_{\{s\}} exp\left( K \sum_{<i,j>} s_i s_j \right) = \sum_{\{s\}} \prod_{<i,j>} exp\left(K s_i s_j\right) \tag{3}$$

where $K = J/(k_B T)$. Note that for simplicity, only the interactions $J$ between the nearest neighboring spins are considered in the 3D Ising model. The relation between the dual lattices is identified:

$$K^* = -\frac{1}{2} ln(tanhK) \tag{4}$$

Clearly, it is just the Kramers-Wannier relation for the definition of $K^*$ in [24,26-29,51,52], for instance, see Eq. (14) in Onsager's work [26]:

$$K^* = \frac{1}{2} ln(cothK) = tanh^{-1}(e^{-2K}) \tag{5}$$

We also have the following identities [26]:

$$tanhK \equiv e^{-2K^*} \tag{6}$$

$$tanhK^* \equiv e^{-2K}$$

(7)

$$sinh2Ksinh2K^{*} = cosh2Ktanh2K^{*} = tanh2Kcosh2K^{*} \equiv 1$$

(8)

It is clear that the physical properties of the 3D $Z_2$ lattice gauge theory with interaction $J^*$ can be determined by those of the 3D Ising model on its dual lattice with interaction $J$. The mapping between the 3D Ising model with the two-spin interaction $J$ and the 3D $Z_2$ lattice gauge theory with the four-spin interaction $J^*$ can be illustrated for simplicity in Figure 4.

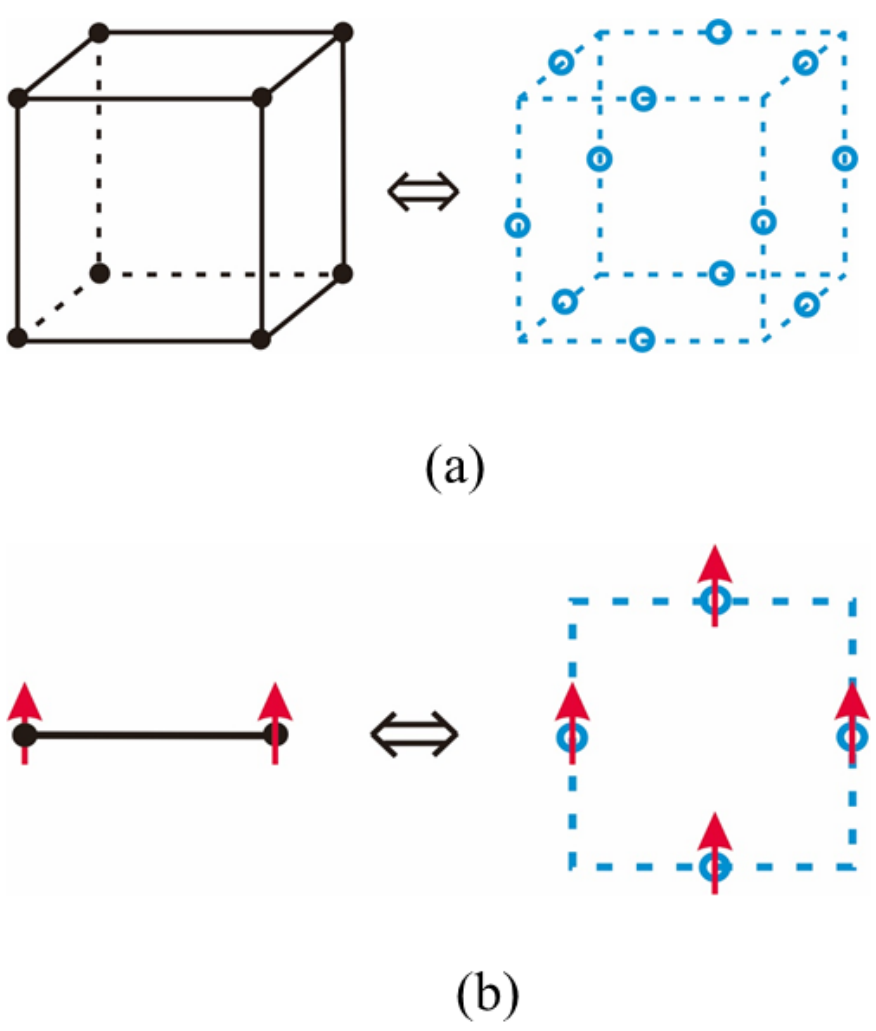

Figure 4. (a) Mapping between the 3D Ising lattice and the 3D lattice of the $Z_2$ gauge theory. (b) Mapping between a two-spin interaction for a link in the 3D Ising model and a four-spin interaction for a plaquette in the 3D $Z_2$ lattice gauge model.

The partition function of the 3D $Z_2$ lattice gauge theory is represented in Eq. (2), in which the summation of spin states takes place for all the possible combination of four spins $\sigma_3(n,\mu)\sigma_3(n+\mu,\nu)\sigma_3(n+\mu+\nu,-\mu)\sigma_3(n+\nu,-\nu)$ along every plaquette. For the 2D $Z_2$ lattice gauge theory, being a special case of the 3D $Z_2$ lattice gauge theory, it is a self-mapping model of the 2D Ising model. The partition function of the 2D $Z_2$ lattice gauge theory is represented in the formulas as the transfer matrices $\mathbf{V_1}$ and $\mathbf{V_2}$ in [43] (however, with the product j running from 1 to n) for the 2D Ising model. However, for the 3D $Z_2$ lattice gauge theory, the situation becomes much complicated. For the description of the transfer matrices with the products running j, we shall map the notation of positions $(n,\mu),(n+\mu,\nu),(n+\mu+\nu,-\mu),(n+\nu,-\nu)$ for four spins in a plaquette, which are involved in a four-spin interaction, into the running number j. Here the total number of spins located at the edges of the original lattice (see Figure 3) is double to be 2mnl. Clearly, the four-spin interaction involves the long-range spin entanglements or the nontrivial topological structures, which implies the complicated structure of the transfer matrices with non-diagonal elements (needed an additional rotation for diagonalization). Fortunately, according the duality of the two dual lattices illustrated in Figure 3 (and also Figure 4), we can use the partition function of the 3D Ising model to describe the 3D $Z_2$ lattice gauge model, with applying the mapping between interactions $K$ and $K^*$ (Eq. (4)). For detailed descriptions of the mapping between the 3D Ising model and the 3D $Z_2$ lattice gauge model, readers refer to subsection 3.A of Wegner's article [18], subsection V.E of Kogut' review article [21], and subsection II.B of Savit's review article [24]. As described in Eq. (2.20) of ref. [24], the partition function of the 3D Ising model can be written in terms of the products of $C_k(\beta)$ and $\left(s_i s_j\right)^k$, while $C_k(\beta)$ is defined as $C_0(\beta)=cosh\beta$ and $C_1(\beta) = sinh\beta$. Collecting together all factors of each spin $s_i$, the partition function can be expressed in terms of the product of a product of $C_{k_\mu}(\beta)$ over *l* and a product of $(s_i)^{\Sigma_i k_\mu}$ over *i*. Considering two possible values (+1 and -1) of the Ising spin, the partition function can be formulated in terms of the product of a product of $C_{k_\mu}(\beta)$ and a product of $2\delta_2\Sigma_i k_\mu$ (see Eq. (2.21) of ref. [24]). In the mapping procedures above, the product over *l* (links) is a product of all $C_{k_\mu}(\beta)$, there being one $C_{k_\mu}(\beta)$ associated with each link of the lattice. The product over *i* (sites) is a product over all the sites of the lattice. For the 3D Ising model, since there is one $k_\mu$ for each link of the lattice, $k_\mu$ is a three vector, while μ is running over the three lattice directions. The sum $\Sigma_i k_\mu$ denotes a sum over the six *k*'s associated with the six links impinging on each site *i* of the lattice. $\delta_2$ is a Kronecker delta function mod 2: it is zero if *n* is odd and one if *n* is even. In order to satisfy the $\delta_2$ function, we have to design a dual lattice and associate with each link of the dual lattice a variable $A_{\mu;i}$ which takes values ±1. As shown in Eq. (2.22) of ref. [24], $k_{\mu;i}$ is related with the product of four variables $A_{\mu;i}$ associated with the four dual links which border the dual plaquette through which the original link with which $k_{\mu;i}$ is associated passes. Then, the partition function of the 3D Ising model can be written in terms of products of the four variables $A_{\mu;i}$ over all plaquettes on the dual lattice (see Eq. (2.23) of ref. [24]). The dual transformation between the 3D Ising model and the 3D $Z_2$ lattice gauge model is rigorous and well-defined over all configurations, also in the thermodynamic limit.

It is clear that the nontrivial topological structures do exist in the 3D $Z_2$ lattice gauge model. Thus, the exact solution of the 3D $Z_2$ lattice gauge model has the same formulas (Eqs. (9)-(11)) as those for the 3D Ising model (however, using the mapping). Figure 5 illustrates a 3D $Z_2$ lattice gauge model on a 3×3×3 lattice, for an example, which is mapped into the 2D spin arrangement on a (3×3+3×3+3×3) lattice, as in the transfer matrix. Green, purple, blue colors represent the four-spin interactions along each plaquette in three crystallographic directions. It is noticed that the spins are located at the edges of the lattice. Again, the different colors are utilized to illustrate the topological structures. In the transfer matrix, the interaction with blue and green colors show the crossings, indicating the existence of non-trivial topological structures in the 3D $Z_2$ lattice gauge model. According to the duality illustrated in Figures 2-4, the 3D $Z_2$ lattice gauge model (shown in Figure 5) is equivalent to the 3D Ising model with interaction $K^*$ on its dual lattice.

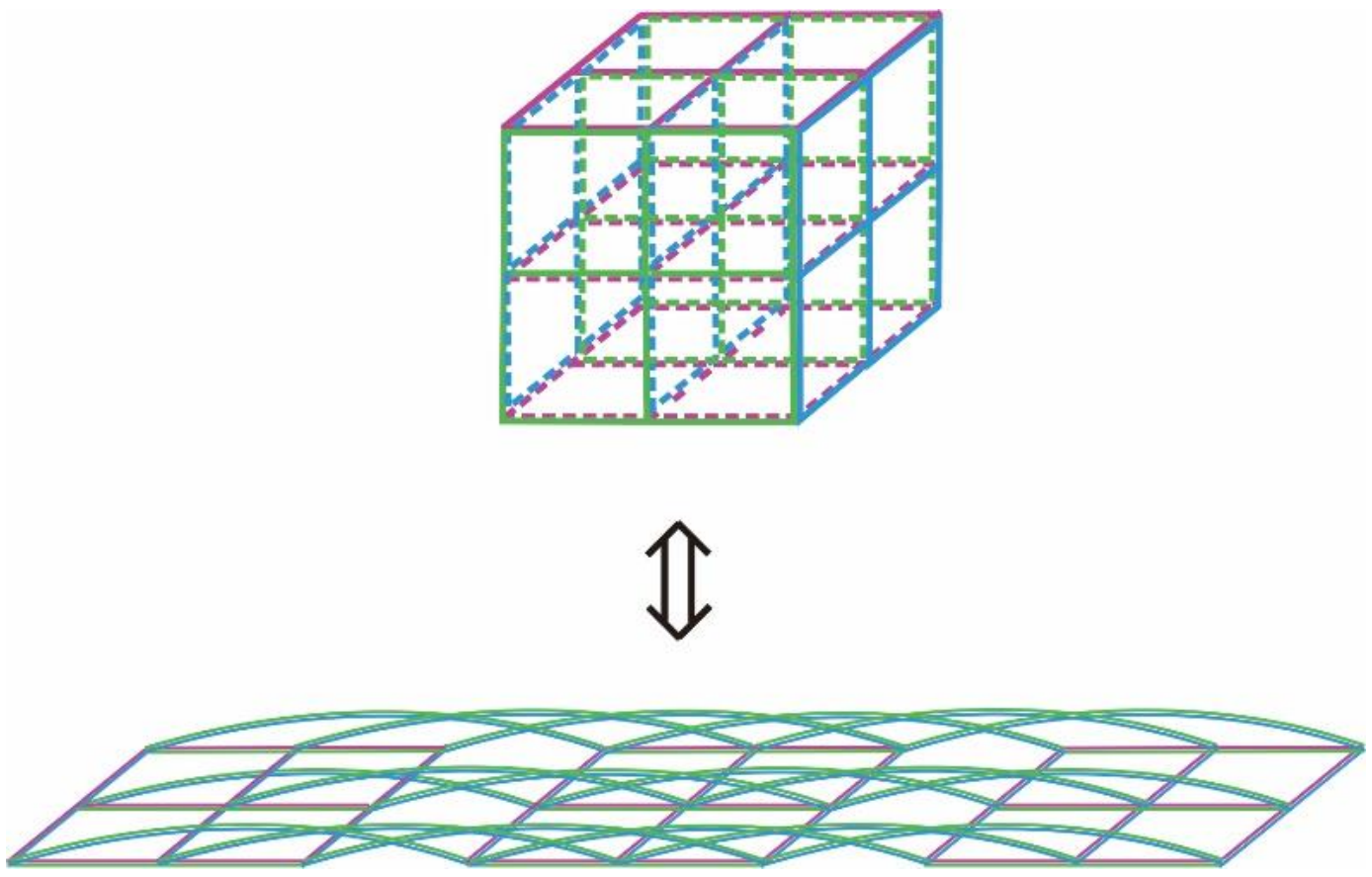

Figure 5. Illustration of a 3D $Z_2$ lattice gauge model on a 3×3×3 lattice, for an example, which is mapped into the 2D spin arrangement on a (3×3+3×3+3×3) lattice, as in the transfer matrix. Green, purple, blue colors represent the four-spin interactions along each plaquette in three crystallographic directions. In the transfer matrix, the interaction with blue and green colors show the crossings, indicating the existence of non-trivial

topological structures in the 3D $Z_2$ lattice gauge model.

It should be noticed that the transformation between the 3D $Z_2$ lattice gauge theory and the 3D Ising model was performed in Eqs. (5.47) - (5.66) of Kogut's review article [21] under a transverse field, so the spins in both the models are quantum spins. If the field strength $\lambda$ were zero, the transformation in Eq.(5.47), Eq.(5.65) and Eq.(5.66) of Kogut's review article [21] could not be proceeded. However, one may first assume the existence of an infinitesimal transverse field, while performing the mapping, them neglecting the effect of the transverse field to maintain the validity of the mapping. Nevertheless, the dual transformation in Eqs. (3.4) – (3.15) of Wegner's paper [18] and Eqs. (2.20) – (2.26) and Eqs. (2.27) – (2.31) of Savit's review article [24] is held without the application of a transverse field, which are classical spin systems. It is concluded that the 3D $Z_2$ lattice gauge theory duals with the 3D Ising model. Therefore, we can utilize the solution of the ferromagnetic 3D Ising model obtained in [27] to derive the exact solution of the 3D $Z_2$ lattice gauge theory, by employing the duality between the two models as provided in [18-21,24]. The ground state, the partition function, the critical point, the specific heat, the spontaneous magnetization, the susceptibility, the spin correlation of the 3D $Z_2$ lattice gauge theory are equivalent to those of the 3D Ising model on its dual lattice, which are explicitly derived in [27-29], with a mapping shown in Eq. (4) or (5).

The partition function of the 3D $Z_2$ lattice gauge theory with interaction $J^*$ on a cubic lattice is represented as [27,28,31]:

$$N^{-1} \ln Z = \ln 2 + \frac{1}{2(2\pi)^4} \int_{-\pi}^{\pi} \int_{-\pi}^{\pi} \int_{-\pi}^{\pi} \int_{-\pi}^{\pi} \ln[\cosh 2\,K \cosh 6\,K - \sinh 2\,K \cos \omega'$$

$$- \sinh 6\,\mathrm{K} \times \left[ cos\omega_x cos\phi_x + cos\omega_y cos\phi_y + cos\omega_z cos\phi_z \right] \Big]$$

$$\mathrm{d}\omega' \mathrm{d}\omega_{\mathrm{x}} \mathrm{d}\omega_{\mathrm{y}} \mathrm{d}\omega_{\mathrm{z}}$$

(9)

with a mapping of $K = -\frac{1}{2} ln(tanhK^*)$. The topological phases $\phi_x$, $\phi_y$, and $\phi_z$ at finite temperature equal to $2\pi$, $\pi/2$ and $\pi/2$, respectively [29,33].

The spontaneous magnetization of the 3D $Z_2$ lattice gauge theory with interaction $J^*$ on a cubic lattice is represented as [27]:

$$\mathrm{M} = \left[ 1 - \frac{16\mathrm{x}^8}{(1-\mathrm{x}^2)^2(1-\mathrm{x}^6)^2} \right]^{\frac{3}{8}}$$

(10)

with $\mathrm{x} = \tanh\mathrm{K}^*$.

The true range $\kappa_x$ of the correlation is obtained for the 3D $Z_2$ lattice gauge theory on the cubic lattice [27]:

$$[\kappa_x a]^{3/2} = 2(K^* - 3K)$$

(11)

with $\kappa_x = 1/\xi$, $\xi$ is the correlation length. At the Curie temperature $1/K^*_c$, $\kappa_x \to 0$ or $\xi \to \infty$,

For the 3D $Z_2$ lattice gauge theory on its dual cubic lattice, we have to perform the same process of the local transformation with an angle K, as what we do for the 3D Ising model. The critical point of the 3D $Z_2$ cubic lattice gauge theory is determined also by $K^* = 3K$, $x_c^* = e^{-2K_c^*} = \left(\frac{\sqrt{5}-1}{2}\right)^3 = 0.23606797......$, thus $K^*_c = 3K_c = 0.72181773....$, $1/K^*_c = 1.38539128....$It is easy to check that $x_c = tanhK_c^*$ and $x_c^* = tanhK_c$. Note that in the ferromagnetic 3D Ising model, one has a disorder phase above the critical point $K_c$, a ferromagnetic ordering phase below the critical point $K_c$; in the 3D $Z_2$ lattice gauge theory, one has a phase transition from a weak-coupling deconfining phase to a strong coupling confining phase at the critical point $K^*_c$ [53]. The critical exponents of the 3D $Z_2$ lattice gauge theory are in the same universality class as the 3D Ising model, which are $\alpha = 0$, $\beta = 3/8$, $\gamma = 5/4$, $\delta = 13/3$, $\eta = 1/8$ and $\nu = 2/3$, satisfying the scaling laws [27].

## 4.Comparison with other approximation methods

The critical point and the critical exponents can be obtained by various approximation methods such as, conventional low-and high- temperature expansions, Monte Carlo simulations, renormalization group field theory, tensor network, conformal bootstrap, etc.. The critical point well established by these methods [54] occurs at $K_c = 0.221655(5)$, i.e., $1/K_c = 4.511505(5)$ for the 3D Ising cubic lattice, that is higher than our exact solution $1/K_c = 4.15617384$……The critical exponents obtained by the Monte Carlo renormalization group [55] are $\alpha = 0.110$, $\beta = 0.3265$, $\gamma = 1.2372$, $\delta = 4.789$, $\eta = 0.0364$ and $\nu = 0.6301$. We need to discuss the possible reasons that cause the differences between our exact solution [27-29] and the approximate values obtained by various methods, widely accepted by the community.

In general, one meets serious hinders at an early stage as one attempts to apply the algebraic method used for solving the 2D model to the 3D model. It is not only because the operators of interest generate a much large Lie algebra that it would be of little value [56], but also because the nontrivial topological structures exist as revealed in Section

2. Indeed, the combinatorial method of Kac and Ward [57] introduces some troubles in topology, which cannot be generated in any obvious way for the 3D problem by counting closed graphs. The peculiar topological property is that a polygon in three dimensions does not divide the space into an "inside and outside" [56]. Any approaches based on only local environments cannot be exact for the 3D Ising model, even though they may be exact for the 2D cases [27,28]. We will discuss the disadvantages of several widely used techniques as follows (for details, the readers refer to [27,28,36,37]):

4.1 Low-temperature series expansions

In principle, all systematic methods for determining series coefficients are at some level graphical or diagrammatic. One simply sums all contributions to calculate the required coefficient. For conventional low-temperature expansions, one chooses the fully aligned spin-up state as the ground state having zero energy to define the partition function, since the low-temperature series is a perturbation expansion about this state [56,58-60]. The conventional low-temperature series expansions consider only the local environment of spins and their interactions, as evaluated by systematically overturning spins from the ground state, neglecting the nonlocal behavior of the 3D Ising model (i.e., the long-range spin entanglements in the plane). Therefore, the conventional low-temperature expansions account only the local part of the partition function (as well as the free energy) of the 3D Ising system. It gives the same fundamental leading term for the 2D triangular Ising lattice and the 3D simple cubic lattice. This is because accounting the energy change of spin flipping, it relates directly to the coordination number q of the Ising model lattice, regardless of the spatial dimensions. Lacking the nonlocal effects causes the divergence of the low-temperature series expansions for the 3D Ising model. In our procedure for deriving the exact solution, the introduction of the fourth dimension realizes its contribution on the spontaneous magnetization, while removing instantaneously the trouble of the dominant singularity on the negative real axis, leading to the additional contribution of the free energy due to the 3D topologic problem.

4.2 High-temperature series expansions

The conventional high-temperature series expansion is an exact expansion for the 2D Ising model and one can easily write down every term of the expansions, by accounting for the loops (polygons) in the 2D lattice. The loops are defined as the boundary between the areas of spin up and spin down [56,58-60]. However, one meets the challenge to write down every term of the high - temperature series expansion for the 3D Ising model, due to the fact that in 3D, the boundaries between domains of spin up and spin down are not polygons, but polyhedra. The polygons used for the 2D case cannot separate the domains of spin up and spin down in the 3D lattice. Furthermore, the existence of nontrivial topological objects (such as crossings of knots/links/braids) in the 3D Ising lattice makes the topologic problems. Actually, any polygons are local, which cannot take into account for the nonlocal behavior of the 3D Ising system, which involves all the states of entangled *mn* spins in a plane (for an interaction between the neighboring spins along the third dimension). According to the duality between the high-temperature and low-temperature expansions, the divergence of the conventional low-temperature expansions corresponds to the zero radius of the convergence of the conventional high–temperature series. It is known that the radius of the circle of convergence for high-temperature expansions has not been proven rigorously yet [31,61,62]. A topological phase transition occurs at/near infinite temperature for the 3D Ising model [27,43], where the trivial topological structure at infinite temperature ($T = \infty$) is changed to the nontrivial topological structure below infinite temperature ($T < \infty^{-}$). It provides the possibilities of several different high-temperature expansions. Namely, the high-temperature expansions are not unique, and the conventional high-temperature series expansion is appropriate for the topological phase transition region ($T = \infty - \infty^{-}$), the correct one at finite temperature ($T < \infty^{-}$) should be the exact solution including the nonlocal behavior [27].

4.3 Renormalization group

Although the renormalization group theory has been widely applied to study the critical phenomena, and shows high-precision results for 2D Ising model, it is not rigorous [63,64]. During the procedures for 3D Ising model, the renormalization group employ various approximations, such as expansions, perturbations, linearizations, normalizations, etc. Thus it possesses the disadvantages similar to the conventional series expansions. The final results of the real-space renormalization group sensitively rely on the processes of dividing Kadanoff blocks, defining the effective Hamiltonian, determining the details of the block variables, and calculating approximately the partial trace. The final results would approach the exact solution if and only if the size of the Kadanoff blocks were chosen to be infinite and the infinite terms of the expansion were retained. Another serious trouble is that the nonlocal behavior in 3D involves in the long-range entanglements of all the possible states of $nm$ spins in a plane ($n \to \infty$ and $m \to \infty$). Surely, any procedure with finite blocks cut the plane of $nm$ spins with the nonlocal effect. The field theoretical or k–space renormalization group techniques meets the same difficulty that cannot be overcome, since at the depth level a close connection is caused by the same basic idea of renormalization groups with connections between $\phi$ and a set of block variables. In both two categories, performing the renormalization reduces the degrees of freedom, while losing the information of the system.

4.4 Monte Carlo methods

Although Monte Carlo simulations are known to be powerful numerical techniques [54,55,64,65], any computer simulations (including Monte Carlo method) for the 3D Ising model are limited by the size effects since the number of the configurations increases in a fashion of $2^N$, with the number of spins, $N \to \infty$. Usually, the period boundary conditions are used along three crystallographic directions so that the simulation is performed on a small system with finite size. This technique breaks down the long-range entanglement between the two nearest neighboring spins along the third dimension in the 3D Ising model. This nonlocal behavior depends on all the possible configurations $2^{nm}$ of $nm$ spins in a plane, with $m \to \infty$ and $n \to \infty$. It makes the systematic error of the simulation on any finite sub-systems plus the periodic boundary conditions. However, in a recent Monte Carlo simulation [49,50], the critical exponents of the 3D Ising model obtained by taking into account the long-range interactions of spin chains agree well with our exact solution. This indicates that by including the nontrivial topological contribution, the Monte Carlo simulation can reach the correct results.

4.5 Conformal bootstrap,

It was claimed [66] that the conformal bootstrap has reached the highest accuracy for calculating the critical exponents for 3D Ising model and other physical systems. Using convex optimization of c-parameter within the conformal bootstrap approach to the four-point correlation functions reported the estimates of $\eta$ = 0.036 and $\nu$ = 0.629 for the 3D Ising model [66]. The disadvantages of the conformal bootstrap are briefly summarized as follows: Although the conformal bootstrap has the highest precision among all the approximation methods, but the same as others, with a systematical error. The systematical error originates from missing the contribution of the global effect to physical properties. The conformal field theory is not a first principle technique but requires nontrivial theoretical input. The bootstrap in statistics cannot account for the global effect, which is equivalent to the true value of the population parameter in the context of statistical inference. This is because the principle of the bootstrap is to replace the overall by parts, treating the sample data as if it is the population, in order to estimate the overall parameters, obtaining the low bound of the overall standard deviation.

It is not the purpose of this work to investigate in details the disadvantages of all the approximation methods. We only give the further explanation why the multitude of separate determinations of the critical exponents throughout the years, by various independent scientists and using completely different techniques coincide. Superficially, all of these different techniques are independent of each other, but in the deeper level, they are related and connected closely. The key factor is that the systematical errors exist seriously in these approximation techniques, which are caused by the reasons discussed above. The systematical errors of these approximation/perturbation techniques are related directly to the physical conceptions/pictures at the first beginning, which neglect the contributions of the nontrivial topological structures to the partition function, the free energy and the subsequent thermodynamic properties. The systematical errors are intrinsic, which cannot be removed by the efforts of improving technically the precision of these approximation/perturbation techniques.

The approximation methods, such as the renormalization group theory and Monte Carlo simulations, are still powerful techniques for the study of the critical phenomena. As suggested in [37], one may obtain the nonlocal contribution to the partition function and also the thermodynamic properties by comparing the approximations with the exact solution. Thus, the nonlocal part of the physical properties (such as spontaneous magnetization) of the 3D Ising model can be obtained by extracting the approximation results from the exact solution. On the other hand, one may still use these approximation techniques, but focus on the structures illustrated in Figure 5 of ref. [33] (see also Figure 1 in ref. [49]), which consist of two parts of contributions (local spin alignments and nonlocal long-range spin entanglements). The results obtained by these methods for such structures (including the nonlocal effects) would be close to the exact solution.

## 5. Physical significance and mathematical aspects

In what follows, we shall discuss the physical significance and the mathematical aspects of the exact solution of the 3D $Z_2$ lattice gauge theory, which are quite helpful for understanding the phenomena in other physical systems.

5.1 Dimensionality, duality, symmetry, manifold, degenerate states and magnetic field

At first, it is interesting to study in more detail the following fundamental issues:

*Dimensionality*: The dimensionality is one of the most important factors in many-body interacting spin (particle) systems, which crucially determines the topological structures of the systems, thus require the particular mathematical techniques. The 3D $Z_2$ lattice gauge theory (the same as the 3D Ising model) must be dealt with in the (3+1)-dimensional space-time [27-29], within the Jordan-von Neumann-Wigner framework [67] with employing Jordan algebras [68,69], which provides the mathematical basis of quantum mechanics and quantum statistical mechanics [43]. Furthermore, it is necessary to perform a time average of t systems of the 3D $Z_2$ lattice gauge models for calculating the physical properties of the 3D $Z_2$ lattice gauge theory [27-29,43], while considering the geometry of the space-time. This verifies that the ergodic hypothesis working for the equilibrium statistical mechanics, which usually takes only the ensemble average is violated in the 3D $Z_2$ lattice gauge theory as in the 3D Ising model [27-29,43,70].

*Duality:* It has been well accepted that the 2D Ising model is self-dual, and its high-temperature paramagnetic phase is dual to its low-temperature ferromagnetic phase. Onsager utilized the Kramers-Wannier relation [51] for the definition of $K^*$ to derive the exact solution of the 2D Ising model [26]. It should be pointed out that such a duality is not invalidated by different numbers of states of paramagnetic and ferromagnetic phases. The ferromagnetic phase of the 2D Ising model has two degenerate ground states (spin up and spin down), while the paramagnetic phase has a unique ground-state (all spins in a disorder state). However, as detailed in the analysis below, the degenerates of the paramagnetic phase are also two, owing to the $Z_2$ symmetry of spins. Thus, it does not cause any problems for the mapping between the ferromagnetic and paramagnetic phases. A spontaneous symmetrical breaking occurs at the critical point, as soon as the system chooses to align spins pointing to the direction of one of the two degenerate states, the system is de-degenerated. It means that one of two degenerate states in the paramagnetic phase is mapped to one of the two ferromagnetic states. In this work, we have employed a well-accepted mapping between the high- (low-) temperature properties of the 3D $Z_2$ lattice gauge system and the low- (high-) temperature properties of the 3D Ising spin system [18,21,24]. In [71], the solution of the ferromagnetic 2D Ising model with a transverse field is derived by the equivalence between the ferromagnetic 2D transverse-field Ising model and the ferromagnetic 3D Ising model, as developed by M. Suzuki [72] (see also [53]). The 3D $Z_2$ lattice gauge theory can be mapped also to the ferromagnetic 2D Ising model with a transverse field [53]. It is emphasized here that mapping/equivalence between the 3D $Z_2$ lattice gauge system, the 3D Ising spin system and the 2D transverse-field Ising system is exact, as clearly seen from the connection between their Hamiltonians. The number of the ground states does not cause any problems with these mappings, since the degenerates of the ground states of the ferromagnetic/paramagnetic phases are invariance.

The Kramers-Wannier relation [51] was developed originally for solving the2D Ising models. This relation is actually the star-triangular relation that corresponds to the Yang-Baxter equation [28]. In the 3D case, the Yang-Baxter equation is generalized to be the generalized Yang-Baxter equation (so-called tetrahedron equation). The star-triangular relation is a special solution of the tetrahedron equation [28], which is also validated for the 3D Ising model [27] and the 3D $Z_2$ lattice gauge theory. The duality mapping between the partition functions of the 3D Ising model and the 3D $Z_2$ lattice gauge theory is exact as a whole. When one starts the mapping in the local environments of a spin, difference in the degrees of freedom may occur so that some constrained conditions arise. The values ±1 of the Ising spin $s_i$ on the original lattice and the variable $A_{\mu;i}$ with the values ±1 on the dual lattice together satisfy the constrain conditions, removing difference in the degrees of freedom. Thus, the duality mapping between the two models can be defined as we work within the gauge-invariant subspace. The gauge invariance maintains the degrees of freedom unchanged locally and globally, even in the thermodynamic limit. Eqs. (5.54)-(5.57) in Kogut's review [21] are satisfied, and also see Eqs. (2.24), (2.25) and Eq. (2.30) in Savit's review [24] for the gauge invariance and the duality mapping, respectively. Furthermore, Wegner discussed the closure condition and the completeness regarding to the topology of the lattice (Eqs. (3.18)-(3.28) in [18]).

*Symmetry*: The symmetry of many-body interacting spin systems and the symmetry of spins both are very crucial for physical properties. The solutions of the many-body interacting spin (or particle) systems should satisfy the symmetries of the system and the spins (or particles) [27-29,32,33,43]. The Ising spins have $Z_2$ symmetry, while the 3D Ising model possesses the global $Z_2$ symmetry for the physical properties. The 3D $Z_2$ lattice gauge theory has a local $Z_2$ symmetry around an Ising spin, so that a gauge transformation can be performed on all the spins in the system.

There are several phase transitions with symmetrical breaking down in the 3D $Z_2$ lattice gauge theory (or the 3D Ising model). Besides the critical point at finite temperature between the ferromagnetic and paramagnetic phases, a topological phase transition occurs at/near infinite temperature for the 3D Ising model [27-29,43], which can be mapped to a gap at/near zero temperature of the 3D $Z_2$ lattice gauge theory. This is consistent with the third thermodynamic law, which states that the absolutely zero temperature cannot be realized in physical systems (like the 3D $Z_2$ lattice gauge model and the 3D Ising model). Meanwhile, the gap existing at/near the absolutely zero in the 3D Ising model can be mapped into a topological phase transition at/near infinite temperature in the 3D $Z_2$ lattice gauge theory [43]. Besides, there are some connections between the structural symmetry of the models and the solution of the critical point. The solution of the critical point is connected to the most beautiful number in the nature (the golden ratio for the 3D Ising cubic lattice model or its third power for the 3D $Z_2$ cubic lattice gauge theory), indicating that the balance between the interaction and the thermal activity in a many-body interacting physical system with the symmetry for the three spatial directions reaches a critical point of the most beautiful number for an order-disorder phase transition.

*Manifold:* It is worth inspecting manifolds of the 3D Ising model and the 3D $Z_2$

lattice gauge theory. Let us first look at the manifold of the 2D Ising model on a lattice $R^2$. The manifold can be constructed by connecting the boundaries along two crystallographic directions to be a torus $S^1 \times S^1$. This process can be generalized to the 3D cases on a lattice $R^3$, so that the manifold of the 3D Ising model (or the 3D $Z_2$ lattice gauge theory) is $S^1 \times S^1 \times S^1$. However, due to the long-range spin entanglement of internal factors, the nontrivial topological structure exists in the transfer matrices of the 3D Ising model [27-29,32,33,43]. It requires the introduction of an additional dimension to the 3D Ising systems, performing the time average, so that the (3+1)D space-time framework is constructed to form a manifold $S^1 \times S^1 \times S^1 \times R^1$. If we perform the time average from $-\infty$ to $\infty$ (or with a closed path from $-\pi$ to $\pi$) for the monodromy representation, the manifold $S^1 \times S^1 \times S^1 \times R^1$ can be transformed to $S^1 \times S^1 \times S^1 \times S^1$. Trivializing the nontrivial topological structure leads to a trivial topological structure in the partition function, while generating a nontrivial manifold for quaternionic wavefunctions with topological phases $\phi_x$, $\phi_y$, and $\phi_z$ (see the eigenvectors in Eq. (33) and discussion on page 5323 of [27]). Note that x, y, and z here denote three directions i, j, and k for quaternionic spaces of wavefunctions, not for crystallographic directions of the original Ising spin lattice. It is known that the unit quaternions can be thought of as a choice of a group structure on the 3-sphere $S^3$. So, the manifold of the quaternionic wavefunctions can be represented as $S^3 \times R^1$ (or 4-ball). In this way, a nontrivial topological problem in a manifold ($S^1 \times S^1 \times S^1 \times R^1$) can be transformed to a trivial topological problem in a manifold (4-ball) with nontrivial topological phases [27-29,32,33,43].

*Degenerate states:* The degenerate states of many-body interacting spin systems rely on the symmetry of spins and the dimensionality of the system. For the 2D Ising model, it is clearly visible that two degenerate states (all spin up and all spin down) exist in the ferromagnetic phase below the critical point, which correspond to two minima in the free energy. Only one minimum appears in the free energy of the paramagnetic phase above the critical point. However, the paramagnetic phase in which all the spins align disorderly has two degenerates also, because due to the $Z_2$ symmetry of Ising spins, all the spins in a degenerate of the disorder phase always point to opposite directions of the spins in another degenerate of the disorder phase. For the 3D Ising models, the topological phases $\phi_x$, $\phi_y$, and $\phi_z$ generated on quaternionic wavefunctions provide the chance of more degenerate states, a kind of topological degenerates. By geometrical and topological concerns, the topological phases $\phi_x$, $\phi_y$, and $\phi_z$ at finite temperature are determined to be $2\pi$, $\pi/2$ and $\pi/2$, respectively [29,33]. From the symmetry of the system, ($\phi_x$, $\phi_y$, $\phi_z$) may have three possibilities ($2\pi$, $\pi/2$, $\pi/2$), ($\pi/2$, $2\pi$, $\pi/2$), ($\pi/2$, $\pi/2$, $2\pi$) [27]. This does not contribute to degenerates, since as we choose the first dimension for the transfer matrix $\mathbf{V}_1$, or if we choose one of the direction as the z-axis, the order for ($\phi_x$, $\phi_y$, $\phi_z$) is fixed, and we have to choose one possibility from three, say ($2\pi$, $\pi/2$, $\pi/2$) [27]. In addition to the dimensionality of the system, the important factor for degenerates is the symmetry of spins, which results in different rotation directions of the topological phases, namely, different chiralities with respect to the directions of spins. The positive/negative angle of the topological phases may correspond to clockwise/counter-clockwise rotations on the manifold, with respect to the $Z_2$ symmetry of Ising spins, which does not change the partition function and the free energy at the long wavelength limit ($\omega_x = 0,\ \omega_y = 0,\ \omega_z = 0$) since all ($\phi_x$, $\phi_y$, $\phi_z$) are in the cos terms (see Eq. (9) above). The combination of positive/negative signs in ($\pm 2\pi$, $\pm\pi/2$, $\pm\pi/2$) loads to that the number of degenerates of the ground-states in the ferromagnetic/paramagnetic phases of the 3D Ising models (and the 3D $Z_2$ lattice gauge models) is eight.

*Magnetic field:* The dual transformation in Eqs. (3.4) – (3.15) of Wegner's paper [18] was carried out in the presence of an external magnetic field along the Ising spin direction (not a transverse field). The dual transformation in Eqs. (2.20) – (2.26) and Eqs. (2.27) – (2.31) of Savit's review article [24] is held without the application of a magnetic field. So, the 3D Ising model and the 3D $Z_2$ lattice gauge model in both two papers [18,24] are classical spin systems. The classical dual transformations proposed by Wegner [18] have the duality relation between the interaction and the magnetic field (see Eqs. (3.16) and (3.17) in [18]). The classical dual transformations proposed by Savit [24] strictly exclude the influence of external fields.

Up to now, our study has been focused on the 3D Ising model and the 3D $Z_2$ lattice gauge theory in the absence of a magnetic field. The nontrivial topological structure of the 3D Ising models exists in the transfer matrix $\mathbf{V}_3$ in the absence of the external magnetic field, the application of a small magnetic field does not affect this effect evidently. If there is a small external field, its correction of the topological structure can be ignored. But, if a large external field is applied, another type of the nontrivial topological structure will appear in the transfer matrix associated with the magnetic field. The problem becomes much more complicated. The exact solution of the 2D (and also 3D) Ising model with a magnetic field is still an open unsolved problem.

5.2 Superfluid, superconductors, and particle physics

Second, it is worth connecting the present magnetic model to other physical systems.

*Superfluid:* The superfluid is one of the most exciting topics in condensed matter physics [73]. The exchange interactions coupling the spins of electrons localized in an assembly of orbitals, of which a typical pair, can be represented as exchange terms in the Heisenberg Hamiltonian (and also the Ising Hamiltonian) [74]. In certain condition, the Heisenberg (or Ising) Hamiltonian can be mapped to the Hubbard model (or t-J model) for hopping and pairing of electrons [75,76]. The 3D lattice gauge theories can be used to investigate the confining flux tube and the electric flux passing through the link in different fields (such as, superconductors, superfluid, Majorana fermions, etc.) [21,52]. The electric flux on a link is quantized in units of the charge g. Let us focus our interest in the application of the 3D $Z_2$ lattice gauge theory to superfluid. The connection exists between the hopping and pairing model [77] for free spinless fermions and our 3D Ising models [21,52]. First, take the 2D Ising model as an example, the pair amplitudes going counter-clockwise (say) around an elementary plaquette have the ratios 1 : i : -1 : -i where i is the imaginary unit. One has to wrap the square lattice around a torus (doughnut) $S^1 \times S^1$ and thread flux through a hole. The phase diagram breaks into two superfluid phases: weak-pairing and strong-pairing, both of which have a "$p_x + ip_y$"-wave pair wavefunction (if considering chirality, it would be "$p_x \pm ip_y$") and an energy gap in the bulk of the lattice. The flux is quantized in units of ħc/2e where ħc/e is the fundamental flux quantum for fermions with charge e (ħ is Planck's constant). Second, in three dimensions, an elementary plaquette can exist along three crystallographic directions with interactions $J^*$. With the pair amplitudes analogous to the 2D case, we can have also two superfluid phases. As mentioned above, the (3+1)D manifold $S^1 \times S^1 \times S^1 \times S^1$ forms as a result of wrapping the cubic lattice along three dimensions and performing the time average in a close path. Then a nontrivial (3+1)D manifold $S^3 \times R^1$ with topological phases is constructed by the topological transformation. A quaternionic "$p_t$ + $ip_x$ + $jp_y$ + $kp_z$"-wave pair wavefunction is constructed for two superfluid phases. Here, i, j, and k are the imaginary units, while x, y, and z denote the directions of quaternionic wavefunction. This situation is similar to the band structure of a 3D material, which is determined by three wave-vectors $k_x$, $k_y$ and $k_z$ along three momentum axes (see page 5322 of [27] for a detailed discussion). The combination of positive/negative signs of ($\pm p_x$, $\pm p_y$, $\pm p_z$) (± denotes chirality) again loads to eight degenerate states of the 3D Ising models (and the 3D $Z_2$ lattice gauge models). What we obtained above is a complete set of chiral p-wave pair wavefunction with orbital angular momentum (l = 0, 1) eigenstates. In a superfluid phase in 3D, flux is also quantized, which threads through three holes. This fact is consistent with the quaternionic wavefunction with topological phases. Eight states can be labelled as (0,0,0), (0,0,1), (0,1,0), (1,0,0), (1,1,0), (1,0,1), (0,1,1), (1,1,1) with the entries labelling the parity (even versus odd) of the number of flux quantum going through each of the three holes in the (3+1)D manifold $S^1 \times S^1 \times S^1 \times R^1$. The number of degenerates of the ground-states in the 3D superfluid model agrees with that of the 3D Ising models derived above. It is interesting to remark that in 2D vortices exist, while in 3D more complicated topological structures may exist, which are composited of vortices in three directions.

*Superconductors:* The mechanisms of superconductors, specially, high-$T_c$ superconductors, are one of the most important topics in condensed matter physics [75-80]. The results obtained for superfluid above are also appropriate for superconductors, which provide some new ideas for understanding superconductors. Usually, superconductors are catalogued to be three types: s-wave, p-wave and d-wave. The p-wave superconductors have been descripted by the "$p_x \pm ip_y$" wave pair wavefunction, which has been thought to be suitable for high-$T_c$ superconductors. Since one believes that the superconducting (transport of paring electrons) is limited in a 2D layer in the high-$T_c$ superconductors, the models used to understand their mechanisms are limited to be the 2D models. However, although the interaction between the 2D layers is very weak, it cannot be neglected. As long as the interaction $K''$ along the third direction is unequal to zero, the system should be treated as a 3D system. The high-$T_c$ superconductors should be investigated in the pairing models in the 3D space, in which the nontrivial topological contribution from bulk to physical properties (such as the energy gap, the critical point) may exist. Thus, the physical model for the high-$T_c$ superconductors should be extended to the (3+1)D framework for Hubbard model [75] (or t-J model [76]), as what we have done for the 3D Ising models. The quaternionic "$p_t \pm ip_x \pm jp_y \pm kp_z$"-wave pair wavefunction are analogous to the p-wave orbitals of electrons in an atom, which have momentums l = 0, 1, with four states (l = 0, m = 0; l = 1, m = 0 and ± 1. Again, there are eight degenerates if we consider the $Z_2$ symmetry of spins or chirality. Therefore, we have to construct the quaternionic chiral "$p_t \pm ip_x \pm jp_y \pm kp_z$"-wave pair wavefunction in quaternionic momentum spaces for p-wave high-$T_c$ superconductors

*Particle physics:* The mathematical structures for the 3D $Z_2$ lattice gauge theory can be generalized to be applicable to other lattice gauge theories and gauge field theories [43]. In particular, the discussion for the duality, manifold, degenerate states, surface modes, and others, are very helpful for understanding other many-body interacting spin (particle) systems, thus for solving the problems of the physics of fundamental interactions in particle physics and high-energy physics [1-24]. For instance, the contribution of nontrivial topological structure to physical properties (such as energy) may be applicable for particle systems, which means an extra energy term [27-29,32,33,43].

The elementary excitations of gauge theories have a string nature. Polyakov suggested a hypothesis [81] that the 3D Ising model can be represented as a fermionic string theory based on Neveu-Schwarz superstrings. In the quantum case, one will face the problem of summing over the surfaces. The simplest model of fluctuating surfaces is the free bosonic string. It is described by the sum over the surfaces (self-intersecting and non-self-intersecting) with statistical weight depending only on the area of the surface. Sedrakyan and Kavalov formulated the sign factor in the 3D $Z_2$ lattice gauge model based on a Dirac theory induced on embedded 2D surfaces [82-84]. The fermionic string can be represented as a line, where spins are attached to every point and interact locally with each other [82]. Hence, the dynamics of the strings reduces to the maps of contours to the so-called loop groups, in terms of which it is tempting to write a dynamical equation. On a level of the partition function, the problem is reduced to the sum of random surfaces where spin matter is living. In the 2D Ising model, the same sign-factor leads to the counting of self-intersections of passes, which eventually leads to the appearance of fermions in the problem. The sign factor, called the Kac-Ward factor, is equivalent to the Pauli principle of fermions. In the gauge formulation of the 3D Ising model (namely, the 3D $Z_2$ lattice gauge theory), the weights of surfaces appear to depend on whether surfaces have self-intersection lines or not. Some configurations appear with plus, and others have minus signs, which should be considered in a partition function. But, in the original formulation of the 3D Ising model, such a sign factor does not appear, instead, the internal factor appears in the transfer matrices. Both the sign factor in the 3D $Z_2$ lattice gauge theory and the internal factor in the 3D Ising model represent the nontrivial topological structures in the systems, but with different degrees of freedom in the local environments, which must satisfy some constrain conditions. It should be emphasized that in the 3D space, the Wilson loop of the gauge theory obeys the area law rather than the perimeter law. This change does not affect the mapping relationship between the two present models.

### 5.3 Topology, geometry and algebra

Third, it is important to point out that the 3D Ising models can serve as a platform for the interplay between the physical properties of many-body interacting systems, topology, geometry, and algebra.

*Topology*: The nontrivial topological structure indeed exists in the 3D $Z_2$ lattice gauge theory, caused by the long-range entanglement between spins, even if only the nearest-neighboring spin interactions are considered in the system [27-29,43,70]. Performing the topological transformation [40-42] as well as the gauge transformation [1,20,21,24] to trivialize the nontrivial topological structure adds an additional contribution to the partition function, the free energy and other thermodynamic physical properties of the 3D $Z_2$ lattice gauge theory. The topological contribution to the physical properties can be connected to the Jones polynomial [85] with the formulas of Wilson loop [19,86] and Witten integral [87] for the action of the gauge group. This indicates that any approaches concerning only the local environments of spins cannot give a correct answer to the desired solution.

*Geometry*: The geometrical structure of the 3D $Z_2$ lattice gauge theory should be described as geometric relations in a hyperbolic 3-sphere (or 4-ball) represented in a (3+1)-dimensional Poincaré ball model [27-29], which is an extension of a hyperbolic triangle in the 2D Poincaré disk model for the 2D Ising system [26]. The topological/geometric phase factors are generated on the quaternionic eigenvectors (and the eigenvalues) of the 3D $Z_2$ lattice gauge theory, which are significant analogous to the phase factors in the Aharonov-Bohm effect, the Berry phase effect and quantum Hall effect, etc. [86,88-90]. The topological phases observed in the 3D Ising model and the 3D $Z_2$ lattice gauge theory originate from the gauge transformation in a 3D many-body interacting spin (or particle) system, again being connected with the Jones polynomial [85], the formulas of Wilson loop [19,86] and Witten integral [87].

*Algebra:* Besides Jordan algebras [67-69], the use of Clifford algebra representation is important to uncover the long-range spin entanglement and the nontrivial topologic structure of the 3D $Z_2$ lattice gauge theory [29,36,37]. Such a nontrivial topologic structure does exist also in the local spin language, but it is not easy to see evidently the global effect in the local spin representation. It is important to construct quaternionic eigenvectors for the 3D $Z_2$ lattice gauge theory, since in this way, the quaternionic geometric phases appear in quaternionic Hilbert space as a result of the topological transformation and the monodromy representation [27-29,32,33,43]. The quaternion basis constructed for the 3D $Z_2$ lattice gauge theory represents naturally a rotation in a (3 + 1)– dimensional space-time. The 2D Ising model has a Hilbert space for describing wave functions as a vector. The 3D Ising model originally has a Hilbert space for describing wave functions as two vectors. The nontrivial topological structure in the 3D Ising model intrinsically requires an additional dimension to form quaternions for describing the many-body interacting eigenvectors. Because the original Hilbert space in the 3D model is insufficient, quaternions are needed to expand the calculation. This framework is closely related with several well-developed theories, such as, complexified quaternion [91], quaternionic quantum mechanics [92], and quaternion and special relativity [93].

The 3D $Z_2$ lattice gauge theory (the same as the 3D Ising model) must be dealt with in the (3+1)-dimensional space-time. The introduction of an additional dimension does not change the Hamiltonian of the original model. The nontrivial topological structure hidden in the transfer matrices related to the Hamiltonian require to the introduction of additional dimension to account for its contributions to the physical properties. The topological transformation for trivializing the nontrivial topological structure is actually a Lorentz transformation in the (3+1)-dimensional space-time [28]. The statistical mechanics is a theory for an equilibrium state, but it is for the study of the statistic average of the physical properties in the equilibrium system in the macro levels. In the micro levels, we need to deal with a dynamical process in the (3+1)-dimensional space-time. For the necessity of the (3+1)-dimensional space-time, please refer to our previous work [27-29,32,33,42].

5.4 Application of the duality in computational complexity

Finally, it is worth noting that the duality between the 3D $Z_2$ lattice gauge theory and the 3D Ising model can be generalized to the spin-glass 3D Ising system [94,95]. With the help of the duality, we proved that the spin-glass 3D Ising model can be mapped to a K-SAT problem for $K \geq 4$ in the consideration of random interactions and frustrations. We also proved that the absolute minimum core model in the spin-glass 3D Ising model is equivalent to the K-SAT problem for $K = 3$, and determined the lower bound of the computational complexity of the K-SAT problem (and also the spin-glass 3D Ising model) [94]. We propose here the following strategy for developing an optimum algorithm for calculations of physical properties (such as, the ground state, the free energy, the critical point, the phase transitions and the critical phenomena, etc.) of the three-dimensional spin-glass Ising model (and also Knapsack problems) [95].

1) Fix z-layers ($z = 1, 2, 3, \ldots$) of the absolute minimum core model as an element of

the algorithms, while performing a parallel computation of *l/z* layers of this element.

2) Compare the precision as well as the accuracy of the results obtained by the above procedures, and determine the optimum value of z.

In this way, one can design the optimum algorithm to find/reach the exact solution with the sufficient accuracy and within the high precision in the shortest time. It can be improved greatly from the present status of $O(1.3^N)$ to $O((1 + \varepsilon)^N)$ with $\varepsilon \rightarrow 0$ and $\varepsilon \neq 1/N$ [30,94,95]. The study on the computational complexity and the optimum algorithm can be extended to be applicable for other NP-problems, for instance, traveling salesman problem, knapsack problem, neural networks [96-99], etc.

## 6.Conclusions

In conclusion, the exact solution of the 3D $Z_2$ lattice gauge theory is derived analytically by the duality between the 3D $Z_2$ lattice gauge theory and the 3D Ising model, based on the exact solution of the 3D Ising model, which is conjectured in [27] and proven in [29,32,33]. The partition function, the spontaneous magnetization, the true range of the correlation and the critical point are rigorously derived for the 3D $Z_2$ lattice gauge theory. The critical exponents of the 3D $Z_2$ lattice gauge theory are in the same universality class as the 3D Ising model, which are $\alpha = 0$, $\beta = 3/8$, $\gamma = 5/4$, $\delta = 13/3$, $\eta = 1/8$ and $\nu = 2/3$. The exact solution for the critical exponents of the 3D Ising model agree well with experimental data in various materials [44-48]. The exact solution gives a guide on numerical studies of the 3D $Z_2$ lattice gauge theory. The present results for the 3D $Z_2$ lattice gauge theory are helpful for figuring out the features of the 3D lattice gauge theories with other symmetries, such as U(1), SU(2) and SU(3).

The present work would play an important role acting as a network for strengthening the discipline not only between different fields in physics (such as condescended matter physics and high-energy particle physics, etc.) and but also among physics, mathematics, and computer science.

**Acknowledgements**

This work has been supported by the National Natural Science Foundation of China under grant number 52031014.

**Data availability** Data available upon request from the author.

**Conflict of interest** The author declares that this contribution is no conflict of interest.